\documentclass[times,twocolumn,final]{elsarticle}

\usepackage{medima}
\usepackage{framed,multirow}

\usepackage{amssymb}
\usepackage{latexsym}

\usepackage{url}
\usepackage{xcolor}

\usepackage{hyperref}
\usepackage{amsmath,amssymb,amsfonts}
\usepackage{algorithmic}
\usepackage{graphicx}
\usepackage{textcomp}
\usepackage{subfigure}
\usepackage{multirow}
\usepackage{xcolor}
\usepackage{booktabs}
\usepackage{color, colortbl}
\usepackage{amssymb}
\usepackage{epstopdf}
\usepackage{dblfloatfix}
\definecolor{newcolor}{rgb}{.8,.349,.1}

\journal{Medical Image Analysis}

\begin{document}

\verso{Weibin Yang, Longwei Xu \textit{et~al.}}

\begin{frontmatter}

\title{More Complex Encoder Is Not All You Need}%

\author[1]{Weibin \snm{Yang}\fnref{fn1}}
\author[1]{Longwei \snm{Xu}\fnref{fn1}}
\fntext[fn1]{Weibin Yang and Longwei Xu contributed equally to this article.}
\author[1]{Pengwei \snm{Wang} \corref{cor1}}
\cortext[cor1]{Corresponding author: 
  E-mail address:\href{wangpw@sdu.edu.cn}{wangpw@sdu.edu.cn}}
\author[1]{Dehua \snm{Geng}}
\author[1]{Yusong \snm{Li}}
\author[1]{Mingyuan \snm{Xu}}
\author[1]{Zhiqi \snm{Dong}}

\address[1]{Shandong University, Tsingtao, 266237, ShanDong, China.}

\received{1 May 2013}
\finalform{10 May 2013}
\accepted{13 May 2013}
\availableonline{15 May 2013}
\communicated{S. Sarkar}

\begin{abstract}
U-Net and its variants have been widely used in medical image segmentation. However, most current U-Net variants confine their improvement strategies to building more complex encoder, while leaving the decoder unchanged or adopting a simple symmetric structure. These approaches overlook the true functionality of the decoder: receiving low-resolution feature maps from the encoder and restoring feature map resolution and lost information through upsampling. As a result, the decoder, especially its upsampling component, plays a crucial role in enhancing segmentation outcomes. However, in 3D medical image segmentation, the commonly used transposed convolution can result in visual artifacts. This issue stems from the absence of direct relationship between adjacent pixels in the output feature map. Furthermore, plain encoder has already possessed sufficient feature extraction capability because downsampling operation leads to the gradual expansion of the receptive field, but the loss of information during downsampling process is unignorable. To address the gap in relevant research, we extend our focus beyond the encoder and introduce neU-Net (i.e., not complex encoder U-Net), which incorporates a novel Sub-pixel Convolution for upsampling to construct a powerful decoder. Additionally, we introduce multi-scale wavelet inputs module on the encoder side to provide additional information. Our model design achieves excellent results, surpassing other state-of-the-art methods on both the Synapse and ACDC datasets. 

Code is available at: 
\url{https://github.com/aitechlabcn/neUNet}
\end{abstract}


\end{frontmatter}


\section{Introduction}
\label{sec:introduction}
\begin{figure*}
	\centering
	\includegraphics[width=\textwidth]{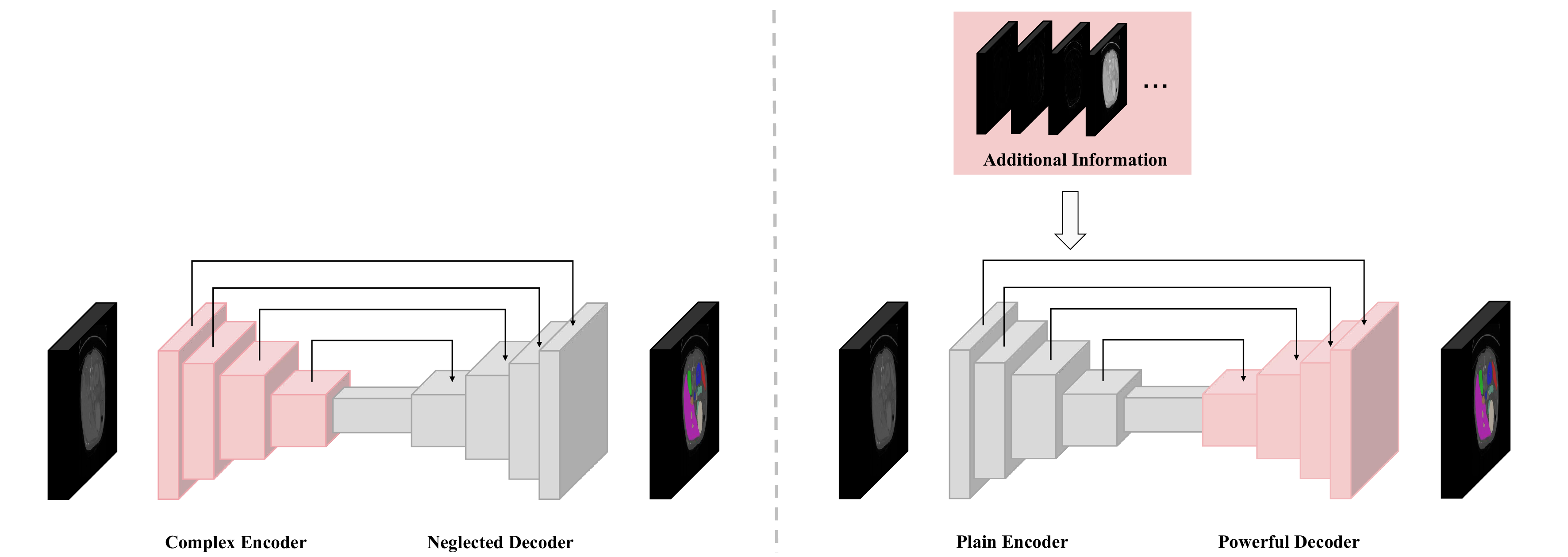}
	\caption{The comparison of our approach (\textbf{right}) with other enhancement methods (\textbf{left}). Currently, the majority of improvement strategies for U-Net aim to construct more complex encoder  to achieve stronger feature extraction capabilities. However, excessively pursuing powerful encoder may not necessarily lead to further improvements in network performance. Therefore, our focus shifts to other aspects of the network, where we endeavor to build more robust decoder part to optimize segmentation details. Meanwhile, We introduce additional information to enhance information utilization efficiency and compensate for information loss.}
\end{figure*}
Image segmentation encompasses the precise categorization of each pixel within an image, constituting a densely predictive undertaking that stands as one of the most pivotal and formidable challenges within the realm of computer vision\citep{review01}. Within the domain of medical image processing, the segmentation of medical images constitutes a pivotal phase within the context of computer-assisted diagnosis.The accurate delimitation of organs or anomalies of interest stands as an indispensable prerequisite for clinical diagnosis.\citep{patil2013medical,norouzi2014medical,elnakib2011medical} Consequently, medical image segmentation has gradually emerged as a focal point within the wider purview of medical image analysis.\citep{pham2000current} \par
The U-shaped Net, known as U-Net\citep{UNet}, is one of the most commonly used networks in medical image segmentation. U-Net employs an Encoder-Decoder network architecture. In this design, the encoder layers are responsible for extracting features from the input image while progressively down-sampling to capture high-dimensional global information\citep{receptive_field}. The decoder, on the other hand, serves two primary functions: (1) gradually upsampling the features to restore the output to the same resolution as the input, and (2) refining segmentation details based on the preceding results. The U-Net encoder-decoder structure is symmetric and incorporates skip connections, which link the feature maps from a specific encoder layer to the corresponding decoder layer. This is done to address the issue of information loss in segmentation tasks while retaining high-resolution features. Due to its excellent segmentation performance, there have been numerous improvements made on the foundation of U-Net in the past.\citep{Attention_UNet,UNet++,UNet3+,swinunet,nnUNet,transunet,zhang2022saa} Most of these efforts did not introduce additional information but focused on designing more complex encoders. For instance, several studies\citep{transunet,swinunet,unetr,swinunetr,nnformer,missformer} introduced self-attention mechanisms and global modeling to achieve more robust feature extraction capabilities.\par 
However, nnU-Net\citep{nnUNet} achieved impressive results without altering the network design, demonstrating that a more complex encoder may not necessarily lead to improved segmentation performance. In successful network designs in the field of deep learning, such as residual connections\citep{resnet}, dense connections\citep{densenet}, and skip connections, the emphasis has been on supplementing additional information rather than creating more intricate encoder designs. Furthermore, there has been limited focus on improving the performance of the decoder in most networks. We believe that both the encoder and decoder have an equally significant impact on the network's results. Without an excellent decoder to progressively restore segmentation maps from high-dimensional abstract features, even the best encoder design may become redundant or inefficient.This paper aims to bridge this gap.
Inspired by the analysis mentioned above, our work primarily proposes improvement strategies in two main aspects: 
\begin{itemize}
  \item[$\bullet$] \textbf{introducing additional information}
  \item[$\bullet$] \textbf{Building a more powerful decoder.}
\end{itemize}
Fig.\ref{sec:introduction} illustrates the distinctions between our improvement strategies and those of prior research. The left section showcases enhancement strategies from previous studies, while the right section presents our strategies. Based on our improvement strategies, with the aim of avoiding the development of a more complex encoder, we have constructed a new network architecture called \textbf{neU-Net} (i.e., \textbf{n}ot complex \textbf{e}ncoder \textbf{U-Net}).

\begin{figure*}
	\centering
	\includegraphics[width=\textwidth]{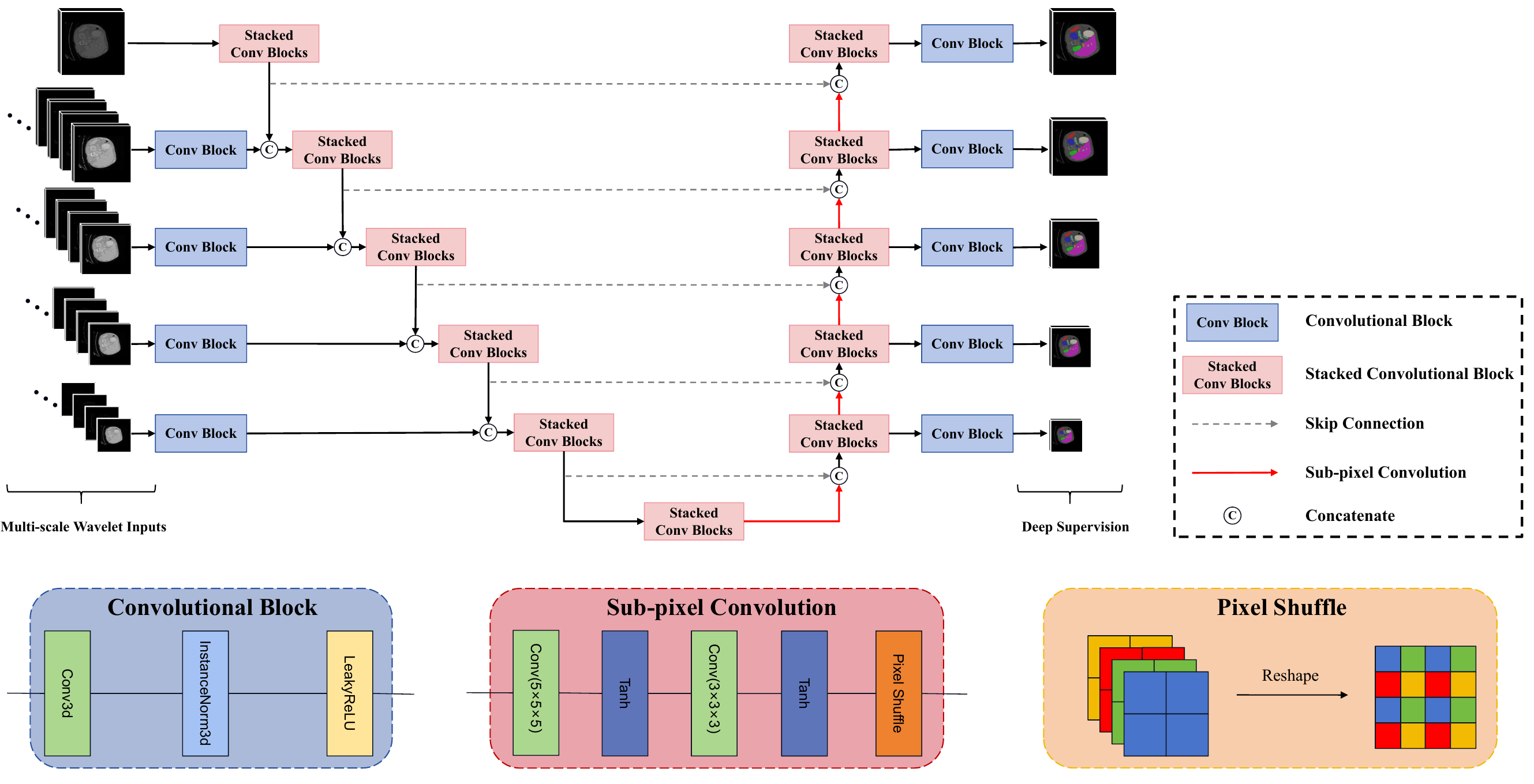}
	\caption{Overview of the neU-Net architecture, On the input side, the input image undergoes wavelet decomposition and is concatenated along the channel, then processed through the convolutional block, which includes sequential 3D convolution, normalization, and nonlinear activation. The output of the block is then concatenated with the output of the encoder from the preceding stage and subsequently fed into the encoder of current layer. In the U-shaped network architecture, Stacked Convolutional Block is composed of two consecutive convolutional blocks. The Sub-pixel Convolution increases the number of feature map's channels through successive convolutions, followed by pixel shuffle to rearrange pixels, thereby achieving upsampling. At the deep supervision layer, the output of each decoder layer is compared with the corresponding downsampled label to compute the loss.}
	\label{fig1}
\end{figure*}

\section{Related work}
In this section, we review U-Net improvement methodologies that are frequently disregarded but significantly contribute to the effectiveness of medical image segmentation, including the introduction of additional information and enhancements to the decoder.

\subsection{Additional Information}
Each module in neural networks can not only receives output feature maps from the preceding module but also has the capacity to incorporate additional information. Additional information provides a richer context, thereby enabling the training of more potent models under constraints of limited data. He et al. introduced the Residual Block \citep{resnet}, which effectively mitigates the problem of vanishing/exploding gradients by employing shortcut connections that add the input features to the output features from a stack of weighted layers\citep{vanishing/exploding}. The U-Net architecture\citep{UNet}, on the other hand, facilitates the efficient fusion of multi-level information by utilizing skip connections to transfer low-level spatial features from the encoder to the decoder. While U-Net restricts skip connections to the same-level encoders and decoders, Huang et al. extended this concept with UNet3+\citep{UNet3+}, employing full-scale skip connections to ensure that each decoder layer comprises larger- and same-scale features from encoders along with smaller-scale feature mappings from other decoder. It is worth noting that the loss of information during the downsampling process in U-Net also impacts the performance of the encoder. However, Residual Blocks confine information propagation within a block, while U-Net and UNet3+ emphasize the augmentation of information for the decoder. To address this issue, Abraham et al.\citep{review8} integrated the image pyramid into the U-Net structure, fusing multi-scale input image information within the encoder phase. Nevertheless, building the image pyramid by directly downsampling the original image or employing Gaussian pyramids will lead to information loss\citep{liu2006multiresolution}. As a reversible transformation, wavelet transform can provide a complete image representation. Moreover, wavelet transform possesses excellent time-frequency locality, enabling the capture of image features at varying resolutions across different regions of the image\citep{wavelets}.

\begin{table*}[!t]
\caption{\upshape Comparison of different hyper-parameters in 3D U-Net, nnU-Net and neU-Net}
\centering
\resizebox{\textwidth}{!}{
\begin{tabular}{|c|c|c|c|}
\hline
Parameters& 
3D U-Net& 
nnU-Net& 
neU-Net(ours) \\
\hline
input patch size& 
fixed& 
task-relevant & 
task-relevant and multi-scale wavelet-based \\
\hline
input spacing& 
fixed& 
task-relevant & 
task-relevant \\
\hline
number of network layers& 
5& 
4-7& 
6 \\
\hline
convolution kernel sizes& 
3$\times$3$\times$3& 
3$\times$3$\times$3 or 1$\times$3$\times$3& 
3$\times$3$\times$3 or 1$\times$3$\times$3 \\
\hline
up(down)-sample ratios& 
(2,2,2)& 
(2,2,2) or (1,2,2) & 
(2,2,2) or (1,2,2) \\
\hline
upsampling methods& 
transposed convolution& 
transposed convolution & 
sub-pixel convolution \\
\hline
\end{tabular}
}

\label{table1}
\end{table*}

\subsection{Decoder}
In the U-Net architecture, the encoder progressively aggregates semantic information at the expense of reducing spatial information through downsampling \citep{nnUNet}. Simultaneously, encoder increases the receptive field by progressively decreasing the size of feature maps to capture multi-scale information. For segmentation tasks, spatial information is crucial for capturing fine-grained segmentation details, and the segmentation result should maintain the same spatial resolution as the input image. Therefore, it is essential to restore spatial information and resolution in some manner, a task typically accomplished by the decoder in U-Net. Consequently, optimizing the decoder plays a pivotal role in enhancing segmentation quality. Zhou et al. introduced UNet++ \citep{UNet++}, which embeds U-Nets of varying depths within the network architecture. All U-Nets share a common encoder, and each level of the U-Net has its independent decoder, interconnected through dense skip connections. Okey et al. proposed Attention U-Net \citep{Attention_UNet}, which employs attention gates (AG) to suppress encoder features that are irrelevant to the decoder features, reducing the semantic gap between the encoder and decoder features. Rahman et al. presented the Cascaded Attention-based Decoder (CASCADE) 
\citep{Cascaded_Attention_Decoding}, which aggregates multiple attention modules during the decoder phase, achieving state-of-the-art (SOTA) results on various datasets. However, these approaches tend to overlook the critical role of upsampling in the recovery capability of decoder. Common upsampling methods, such as interpolation algorithms and transposed convolution, have certain issues. Interpolation algorithms\citep{patil2013medical3}, such as nearest-neighbor interpolation, bilinear interpolation, and cubic interpolation, are the most common upsampling methods. However, for medical images with diverse shapes and intricate structures, the simple weighted summation operation of these algorithms results in limited effectiveness. On the other hand, when the stride and kernel size are not appropriately matched, transposed convolution can lead to the checkerboard problem\citep{checkerboard}. The Sub-pixel Convolution technique, as proposed by W. Shi et al. \citep{patil2013medical5}, offers an alternative perspective for upsampling. The algorithm achieves upsampling by expanding the channel dimension of the feature maps through convolution and then performs periodic shuffling of pixels from the channel dimension to the spatial dimension, enhancing the quality of feature map resolution restoration.

\section{Method}
In contrast to other approaches that prioritize building encoders with powerful feature extraction capabilities, we hold the view that an encoder composed of plain convolutions already possesses sufficient feature extraction capabilities to handle medical image segmentation tasks, which typically have a relatively small dataset size. In fact, more complex encoders may even lead to overfitting\citep{overfitting} We analyze the functions of various components within the U-Net architecture and determined that, in contrast to the encoder, the decoder is equally crucial and offers significant room for optimization. The decoder part refines segmentation details based on the output of encoder and progressively restore spatial information and spatial resolution of feature maps, while the quality of upsampling results directly influences the performance of the decoder part. The commonly used transposed convolution in 3D U-Net and its variants often suffer from  checkerboard problem. To address this, we design a novel Sub-pixel Convolution method, which effectively enhances the quality of upsampling. Furthermore, information loss during the downsampling process can also impact the performance of encoders. Inspired by skip connections and image pyramids, we employ 3D discrete wavelet transform and supplement the resulting wavelet pyramid on the input side of network, providing aggregate information for each stage of the encoder.

\subsection{Network Architecture}
 Our approach focuses on components beyond the encoder, leading us to name the network neU-Net. Fig.\ref{fig1} presents the network structure of neU-Net, neU-Net comprises the U-shaped encoder-decoder structures similar to the U-Net\citep{UNet} and incorporates the multi-scale wavelet layer on the input side, which provide comprehensive multi-scale information and frequency domain information of input image to each decoder layer. The decoder layers utilize sub-pixel convolution to achieve upsampling, thereby enhancing the quality of the feature maps after enlarging their dimensions. Furthermore, we build our model based on the nnU-Net\citep{nnUNet}framework, which facilitates the adaptive determination of network hyper-parameters such as kernel size, down-sample and up-sample ratios, and the number of network layers based on dataset attributes and training devices. To ensure a dynamic adaptability to different tasks and improve the transferability of model, we maintain the number of network layers at 6. Table.\ref{table1} provides the comparison of the parameter configurations for neU-Net, nnU-Net, and 3D U-Net\citep{3DUNet}. In the preprocessing stage of nnU-Net, the dimension with the smallest size is placed at the forefront. During convolution and up(down)-sample, the frequency of size changes in this dimension is fewer compared to the other two dimensions. The number of network layers refers to the total count of encoder layers combined with the bottleneck layer.
 
 Stacked Convolutional Block is composed of two convolutional blocks consecutively arranged, both utilizing uniform convolution kernel sizes, which are selected based on the dataset characteristics such as anisotropy as either 3$\times$3$\times$3 or 1$\times$3$\times$3. In the encoder layers, the first convolutional block of stacked convolutional block applies a stride of (2,2,2) or (1,2,2), align with the previos layer's convolution kernel size In order to eliminate the influence of anisotropy as much as possible. While extracting features, this convolutional block accomplishes downsampling through stride convolution. The second convolutional block maintains a stride of (1,1,1). In the decoder stage, both convolutional blocks utilize a stride of (1,1,1), only relying on upsampling to change the spatial sizes of the feature maps.
 
\begin{figure*}[t]
	\centering
	\includegraphics[width=0.75\linewidth]{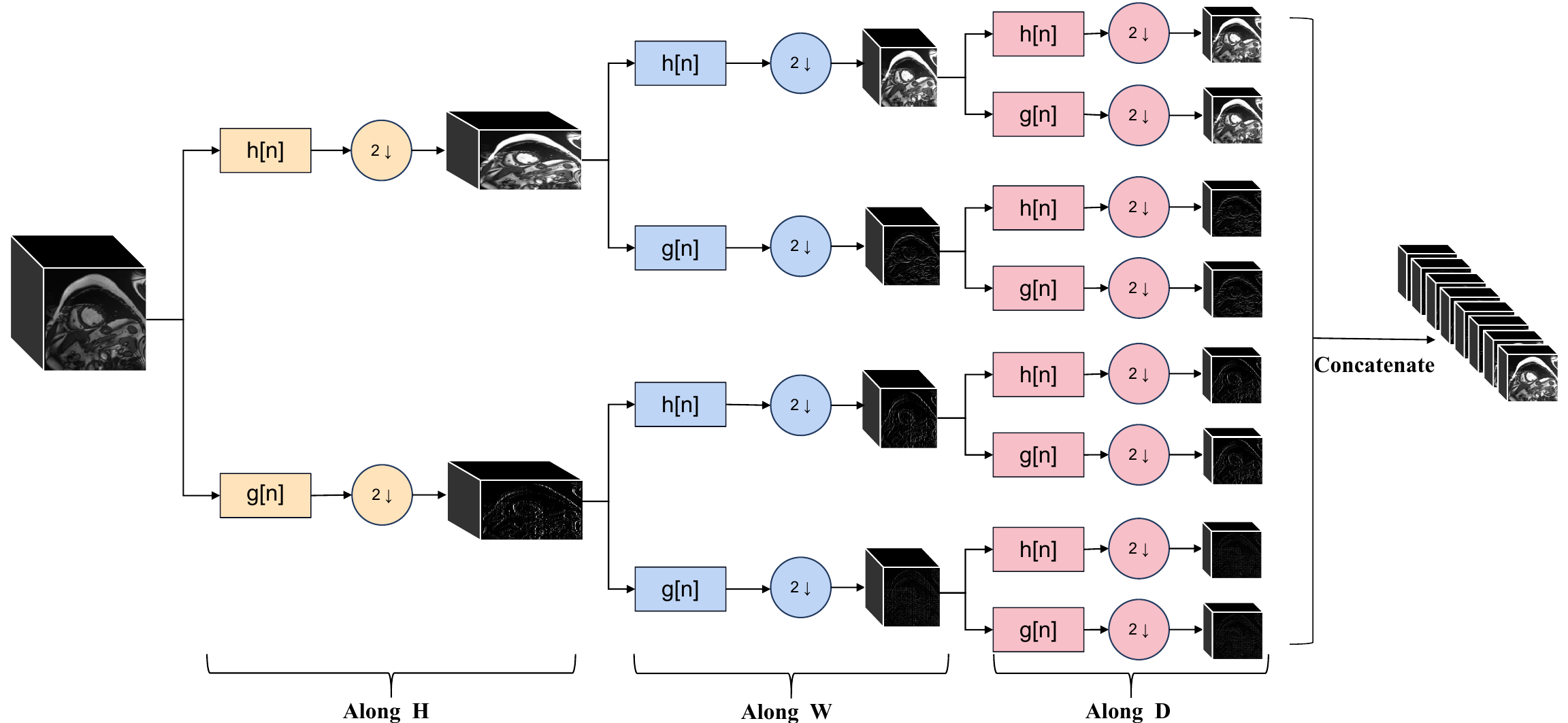}
	\caption{Process of Wavelet Multi-scale Decomposition. In neU-Net, 3D discrete wavelet transform is applied to the input image, guided by the down-sample ratios of the preceding encoder stage. 3D input image is decomposed along its three dimensions, then the decomposed sub-band images are concatenated along the channel dimension. \textbf{h[n]} and \textbf{g[n]} denote the low-pass and high-pass filters respectively, and \textbf{$2 \downarrow$} signifies two-fold down-sampling. Figure illustrates the decomposition process with down-sample ratios of (2,2,2), ultimately yielding 8 sub-bands that represent different features of the input image.}
	\label{fig2}
\end{figure*} 

\subsection{Multi-scale Wavelet Inputs}
To enhance the segmentation accuracy of the network, we have introduced a pyramid-like multi-scale input strategy. In conventional approaches, a common practice is to perform n-fold straightforward subsampling on the original image to obtain multi-scale inputs. However, we posit that such a method does not effectively compensate for the loss of high-frequency information during the continuous downsampling of feature maps.To address this issue, we introduce Discrete Wavelet Transform (DWT) for downsampling, simultaneously preserving low-frequency information and high-frequency edge details in a lossless manner.
The formulation of the one-dimensional Discrete Wavelet Transform is defined as follows, Given an input signal $x[n]$ of length $N$, along with wavelet functions $h[n]$ and $g[n]$ as the decomposition and reconstruction filters, respectively, the computation of DWT involves two steps: low-pass filtering (decomposition) as (\ref{CA}) and high-pass filtering (decomposition) as (\ref{CD}).
\begin{equation}
    cA[k] = \sum^{N-1}_{n=0}x[n]h[2k-n] 
    \label{CA}
\end{equation}
\begin{equation}
    cD[k] = \sum^{N-1}_{n=0}x[n]g[2k-n]
    \label{CD}
\end{equation}
In the above formulas, $k \in \mathbb{R}^{\frac{N}{2}-1}$, $cA[k]$ represents the approximation coefficients, characterizing the low-frequency components of the signal, while $cD[k]$ represents the detail coefficients, capturing the high-frequency components of the signal.We opted to utilize the simplest Haar wavelet for wavelet selection, as it already fulfills our requirements.
In the Haar wavelet, the expressions for the low-pass filter 
$h[n]$ and high-pass filter $g[n]$ are as follows:

\begin{equation}\label{Haarh}
    \begin{split}
        h[n]=\left \{
        \begin{array}{ll}
             \frac{1}{\sqrt{2}}  &n = 0,1 \\
              0                  &otherwise 
        \end{array}
        \right.
    \end{split}
\end{equation}
\begin{equation}\label{harrg}
    \begin{split}
        g[n]=\left \{
        \begin{array}{ll}
             -\frac{1}{\sqrt{2}}  &n = -1\\
              \frac{1}{\sqrt{2}}  &n = 0\\
              0                  &otherwise 
        \end{array}
        \right.
    \end{split}  
\end{equation}

In the context of medical image coordinates,defining the shape of volume as follows: length $H$, width $W$, and depth $D$,Assuming the input volume is denoted as $I(i,j,k)$, where $I \in \mathbb{R}^{^{H\times W \times D}}$. As in Fig.\ref{fig2} we initially apply a one-dimensional Discrete Wavelet Transform (DWT) to the i-axis.
Subsequently, DWT is performed on the resulting two transformed coefficients along the j-axis, yielding four coefficients. Following this, a wavelet transform is applied to the four coefficients along the k-axis, resulting in the final eight coefficients.\par
Defining $cA$ represents the approximation coefficients, $cD_{l}, l\in \left \{1,2,3,4,5,6,7 \right\}\ $ represents the detail coefficients along different directions.$cA$ and $cD_{l}$ are all $\in \mathbb{R}^{^{\frac{H}{2}\times \frac{W}{2}\times \frac{D}{2}}}$.
Upon obtaining the eight coefficients, they are concatenated along the dimensions to generate the $I_{w}$, which represents input result after wavelet transformation.
\begin{equation}
    I_{w} = concatenate((cA,cD_{1}, \dots, cD_{7}), axes=channel)
\end{equation}
where,$I_{w} \in \mathbb{R}^{^{\frac{H}{2}\times\frac{W}{2}\times\frac{D}{2}\times 8}}$.
Starting from the first layer of the encoder, a continuous wavelet transformation is applied to the previous layer's approximation coefficients (with the 0th layer representing the original input volume). This process results in a pyramid-style multi-scale wavelet input.

\subsection{Sub-pxiel Convolution}
Given a input volume $x\in \mathbb{R}^{^{H\times W\times D\times C }}$, where H, W, D and C signify the dimensions of height, width, depth, and channel respectively. The upscaling ratio r can be determined based on the up-sample ratio of the layer. According to Table.\ref{table1}, $r\in \left \{4,8  \right \}$. The volume x undergoes a 5$\times$5$\times$5 convolution and activation through the tanh function, thereby transforming into $x^{'}\in\mathbb{R} ^{H\times W\times D\times 2\cdot C}$, expanding the channel dimension by a factor of two. Subsequently, $x^{'}$ is subjected to a 3$\times$3$\times$3 convolution and activation, outputing $x^{''}\in\mathbb{R} ^{H\times W\times D\times r\cdot C}$. Finally, we use the periodic shuffling operation to reshape the channels of $x^{''}$ to high-resolution output.

\begin{figure}[!t]
    \centering
    \subfigure[sub-pixel convolution]{
    \begin{minipage}[b]{\columnwidth}
        \centering
        \includegraphics[width=\columnwidth]{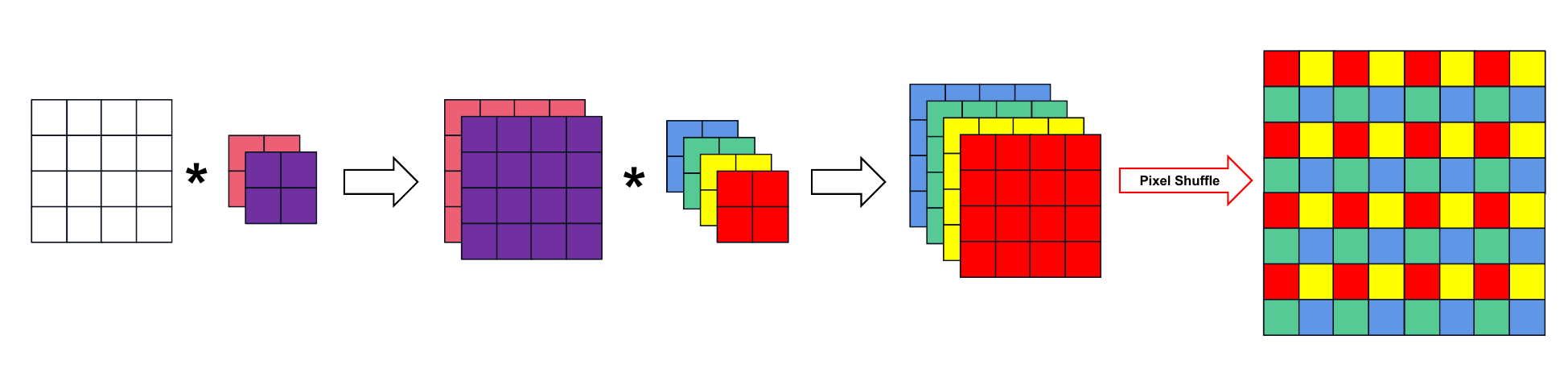}
    \end{minipage}
    }
    \subfigure[transposed convolution]{
    \begin{minipage}[b]{\columnwidth}
        \centering
        \includegraphics[width=\columnwidth]{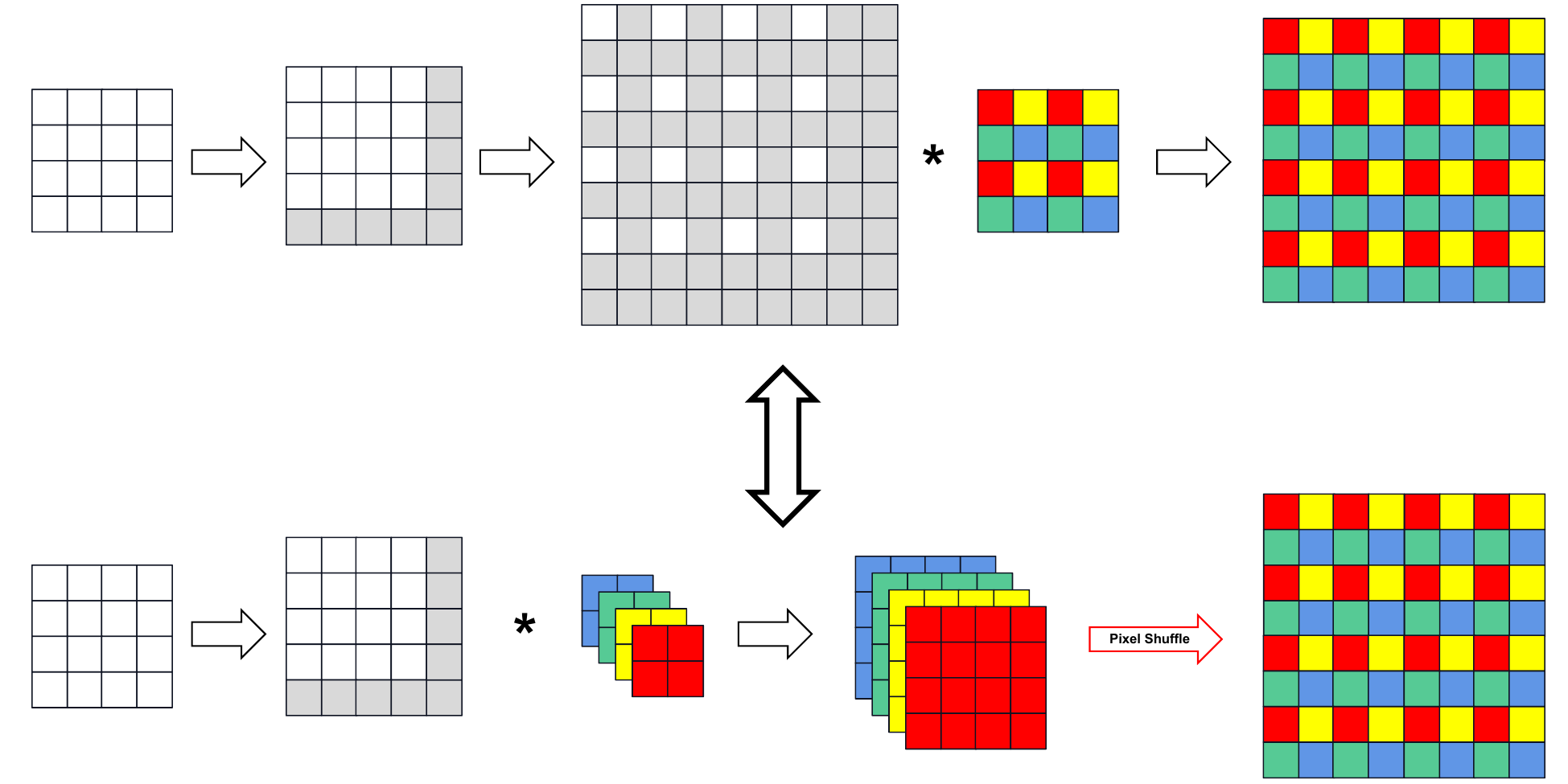}
    \end{minipage}
    }   
    \caption{Calculation Process of Common Upsampling Methods}
    \label{fig3}
\end{figure}

Fig.\ref{fig3} illustrates the process of sub-pixel convolution and transposed convolution for achieving two-fold upsampling of a 4$\times$4 input image, where $\ast$  denotes the convolution operation. For transposed convolution, as shown in the upper part of Fig.\ref{fig3}(b)\citep{patil2013medical4}, the input feature map is initially padded with zero around its sides and between pixels, with gray pixels representing the padding pixels. The padded image is then convolved by a transposed convolutional kernel with stride of 1. It is worth noting that the weights at different positions of the transposed convolution kernel activate independently. For example, the upper-left pixel of the output feature map is activated solely by the red weights. As a result, we can break it into four 2$\times$2 convolution kernels. In this scenario, the process of transpose convolution can be illustrated by the lower part of Fig.\ref{fig3}(b)\citep{transconv2}. Similar to sub-pixel convolution, the four outputing 4$\times$4 feature maps from the transposed convolution are reshaped though the periodic shuffling operation, moving pixels from the channel dimension to the spatial dimension. This also provides an alternative explanation for the checkerboard problem of transposed convolution\citep{PTCN}: the intermediate feature maps are generated by independent convolution kernels, leading to no direct relationship between adjacent pixels on the output feature map. One approach to address the checkerboard problem is using interpolation algorithms during input image padding. However, this method will introduce additional computational overhead.

In contrast, as demonstrated in Fig.\ref{fig3}(a), the sub-pixel convolution algorithm designed by us progressively restores feature map resolution. We increase the number of channels to twice the input channels, then it is further expanded to the complete up-sample ratio. The second convolution layer not only enlarges the feature map channel size further but also amalgamates the features of each output channel from the first convolution layer. This enhances the correlation among adjacent pixels on the output feature map.

\subsection{Loss Function}
We train our networks with a combination of dice\citep{diceloss} and cross-entropy loss, the total loss during the training phase can be formulated as follows:
\begin{equation}
    L_{total} =w_{1} L_{1}+ w_{2} L_{2} + w_{3} L_{3}+ w_{4} L_{4}+ w_{5} L_{5}
\end{equation}
where $L_{i}$, $i\in \left \{1,2,3,4,5  \right \}$ represents the loss of the decoder at the i-th layer. When i equals 1, it represents the topmost decoder layer. Here, $w_{i}$ denotes the weight of loss for the i-th layer of the encoder, the calculation formula is:
\begin{equation}
    w_{i} =\frac{\frac{1}{2^{i-1}}}{\sum_{m=0}^{5}\frac{1}{2^{m}}} 
\end{equation}
The loss for each decoder layer comprises dice loss and cross-entropy loss:
\begin{equation}
    L=L_{dice} + L_{CE}
\end{equation}
The computation formulas for dice loss and cross-entropy loss are as follows:
\begin{equation}
    L_{dice}=1-\frac{2\sum_{c=1}^{C}\sum_{i=1}^{N}g_{i}^{c}s_{i}^{c}}{\sum_{c=1}^{C}\sum_{i=1}^{N}g_{i}^{c}+\sum_{c=1}^{C}\sum_{i=1}^{N}s_{i}^{c}} 
\end{equation}
\begin{equation}
    L_{CE}=- \frac{1}{N} \sum_{c=1}^{C} \sum_{i=1}^{N} g_{i}^{c} \log_{}{s_{i}^{c}} 
\end{equation}
where C represents the number of categories and N represents the number of voxels in 
each category. $g_{i}^{c}$ is the ground truth binary indicator of class label c of voxel i, and $g_{i}^{c}$ is the corresponding segmentation prediction.
\begin{table*}[!ht]
    \begin{center}
    \caption{\upshape Comparison on the abdominal multi-organ Synapse dataset. We use HD95 and DSC to evaluate the performance of each model.The best results are indicated in bold. neU-Net achieved the best performance. Abbreviations stand for: Spl: \textit{spleen}, RKid: \textit{right kidney}, LKid: \textit{left kidney}, Gal: \textit{gallbladder}, Liv: \textit{liver}, Sto: \textit{stomach}, Aor: \textit{aorta}, Pan: \textit{pancreas}.}
        \resizebox{\textwidth}{!}{
        \begin{tabular}{l|c c c c c c c c|cc}
        \toprule
            \multirow{2}{*}{Methods}  & \multirow{2}{*}{Spl} &  \multirow{2}{*}{RKid} &  \multirow{2}{*}{ LKid} & \multirow{2}{*}{Gal}  & \multirow{2}{*}{Liv}  & \multirow{2}{*}{Sto} & \multirow{2}{*}{Aor} &  \multirow{2}{*}{Pan} &  \multicolumn{2}{c}{Average} 
            
            \\ \cmidrule{10-11}
             & & & & & & & & & HD95 $\downarrow$ & DSC $\uparrow$ \\
                \midrule
                \midrule
        U-Net~\citep{UNet} & 86.67 & 68.60 & 77.77& 69.72 & 93.43 & 75.58  &  89.07  & 53.98 & - & 76.85 \\
        TransUNet~\citep{transunet} & 85.08 & 77.02 & 81.87 & 63.16 & 94.08 & 75.62 &  87.23  & 55.86 & 31.69 & 77.49 \\
        
        Swin-UNet~\citep{swinunet} & 90.66 &  79.61 & 83.28 & 66.53 & 94.29 & 76.60 & 85.47 &   56.58  & 21.55 & 79.13   \\
        UNETR~\citep{unetr} & 85.00 & 84.52 & 85.60 & 56.30 & 94.57 & 70.46 & 89.80 & 60.47 &  18.59 & 78.35  \\
        MISSFormer~\citep{missformer} & 91.92 &  82.00 & 85.21 &  68.65 & 94.41 & 80.81 & 86.99 & 65.67 & 18.20 & 81.96 \\
        
        Swin-UNETR~\citep{swinunetr} & 95.37 & 86.26 & \textbf{86.99} & 66.54 & 95.72 & 77.01 & 91.12 &  68.80 &  10.55 & 83.48 \\
        
        nnFormer~\citep{nnformer} & 90.51 & 86.25 & 86.57 & 70.17 & 96.84 & \textbf{86.83}  & 92.04 &    \textbf{83.35} &  10.63 & 86.57\\
        nnU-Net~\citep{nnUNet}  &\textbf{91.86} &88.17 & 85.57 &71.76 &\textbf{97.23} &85.26 &93.01 &83.01 &10.77 &86.98 \\
        \midrule
        \rowcolor{red!6} \textbf{neU-Net(Ours)} & 91.03 & \textbf{89.83} & 85.27 & \textbf{80.89} & 97.20 & 82.82 & \textbf{93.17} & 82.42  &  \textbf{9.13} & \textbf{87.83} \\
        \bottomrule
        \end{tabular}
        }\vspace{-0.5em}
        \label{table:synapse}
        \end{center}
\vspace{-0.8cm}
\end{table*}

\section{Experiments}
\subsection{Datasets}
To validate the effectiveness of our method, we conducted experiments on the Multi Atlas Labeling Beyond The Cranial Vault (BTCV)\citep{BTCV}, Synapse multiorgan segmentation\citep{BTCV}, and Automatic Cardiac Diagnosis Challenge (ACDC) datasets\citep{ACDC}. These datasets encompass different imaging modalities and segmentation tasks, providing a comprehensive evaluation of our model.

\subsubsection{Synapse} Dataset comprises abdominal CT scans from 30 subjects, covering 8 distinct organs: spleen, right kidney, left kidney, gallbladder, liver, stomach, aorta, and pancreas. Following the data split in \citep{transunet}, we select 18 samples for training our model and evaluated it on the remaining 12 samples.

\subsubsection{BTCV} The BTCV dataset consists of 30 training/validation samples and 20 testing samples, with manual annotations conducted under the supervision of radiologists from Vanderbilt University Medical Center. The annotations cover 13 organs, including all 8 organs of Synapse dataset, along with esophagus, inferior vena cava, portal and splenic veins, right adrenal gland, and left adrenal gland. These additional organs include the esophagus, inferior vena cava, portal vein, splenic vein, right adrenal gland, and left adrenal gland. Each CT scan consists of 80 to 225 slices, and each slice having 512$\times$512 pixels with a thickness varying from 1 to 6mm. We select 24 out of the 30 training/validation samples as the training set, and the remaining 6 samples are used as the validation set for conducting ablation experiments.

\subsubsection{ACDC} The ACDC dataset comprises 100 cardiac MRI images, which have been annotated for the left ventricle (LV), right ventricle (RV), and myocardium (Myo). The samples were collected from healthy individuals, patients with myocardial infarction, patients with dilated cardiomyopathy, patients with hypertrophic cardiomyopathy, and patients with right ventricular abnormalities. Following the data split method described in \citep{nnformer}, the dataset was divided into 70 training samples and 10 validation samples, with the remaining 20 samples reserved for testing.

\subsection{Metrics}
We have employed a comprehensive set of two evaluation metrics to rigorously assess the effectiveness of the methodology. These metrics consist of the Dice coefficient, utilized to quantitatively gauge the degree of similarity between the predicted segmentation and the ground truth segmentation. A value converging towards 1 signifies a higher degree of segmentation accuracy. Additionally, we have incorporated the Hausdorff 95 distance, a metric tailored to quantitatively capture the maximum spatial separation between the predicted segmentation and the ground truth. This parameter provides a robust evaluation of the alignment and coherence of segmentation boundaries.The expressions for the two evaluation metrics are provided below:
\begin{equation}
\textrm{Dice} = \frac{2\sum_{i=1}^{I} Y_{i}\hat{Y}_{i} }{\sum_{i=1}^{I}Y_{i}+ \sum_{i=1}^{I}\hat{Y}_{i}},
\label{eq:dice_score}
\end{equation}
\begin{equation}
HD_{95} =\max^{95^{th}} \{{\max _{y' \in Y'} \min _{\Bar{y}' \in \Bar{Y}'} } \|y'-\Bar{y}'\|, 
\max _{\Bar{y}' \in \Bar{Y}'} \min_{y' \in Y'} \|\Bar{y}'-y'\| \}.
\label{eq:hd_score}
\end{equation}
where $Y$ and $\Bar{Y}$ denote the ground truth and prediction of voxel values. $Y'$ and $\hat{Y}'$ denote ground truth and prediction surface point sets. The notation $\max^{95^{th}}(\cdot)$ represents the value obtained by sorting in descending order and selecting the value corresponding to the 95th percentile.

\subsection{Implementation Details}
We implement neU-Net in PyTorch\citep{pytorch} 2.0.0 and nnU-Net 2.1.1. All experiments were conducted on a single NVIDIA GeForce RTX 3090 GPU with 24 GB memory. We follow the default data preprocessing, data augmentation, and training strategies of nnU-Net\citep{nnUNet}. In the data pre-processing stage, we cropped all data to the non-zero regions, then the data will be resampled to the median voxel spacing of the dataset. In the presence of heterogeneous voxel spacings, meaning that the spacing along one axis is three times or more than that of the other axes, the 10 percentile of the spacing will be used as the spatial size for this axis. Finally, the data will be normalized. For CT images, such as BTCV, the intensity values of the foreground portion of the dataset are first collected and the entire dataset is normalized by clipping to the [0.5, 99.5] percentiles of these intensity values. Z-score standard normalization\citep{2.5Dzhangchi} then is applied to the data based on the mean and standard deviation of all the collected intensity values. For MRI images, such as ACDC, or other modalities, individual sample information is collected and z-score normalization is applied to that specific sample. Multiple techniques are employed for data augmentation, including rotation, scaling, Gaussian noise, Gaussian blur, brightness augmentation, contrast adjustment, simulation of low resolution, gamma transformation, and mirror transformation. 
For the Synapse and BTCV datasets, the patch size is set to 48$\times$192$\times$192, with a batch size of 2. As for the ACDC dataset, the patch size is 10$\times$96$\times$96 and the batch size is fixed at 5. We trained our model from scratch with an initial learning rate of 0.01, and updates are performed according to the poly decay strategy:
\begin{equation}
l_{cur}=l_{initial}  \times \left (  1-\frac{E_{cur}}{E_{max}} \right ) ^{0.99} 
\end{equation}
where $l_{cur}$ denotes the learning rate of the current epoch, $E_{cur}$ denotes the number of current epochs, and $E_{max}$ denotes the number of training epochs, which is set to 1000 for Synapse and BTCV, while for ACDC, epochs is set to 400. Furthermore, we employ the SGD optimizer with momentum of 0.99 and a weight decay of 3e-5 to update gradients. We use the Dice Similarity Coefficient (DSC) and the 95\% Hausdorff Distance (HD95) metrics to evaluate our model.
\begin{figure*}[htbp]
 \centering
 \includegraphics[width=0.75\textwidth]{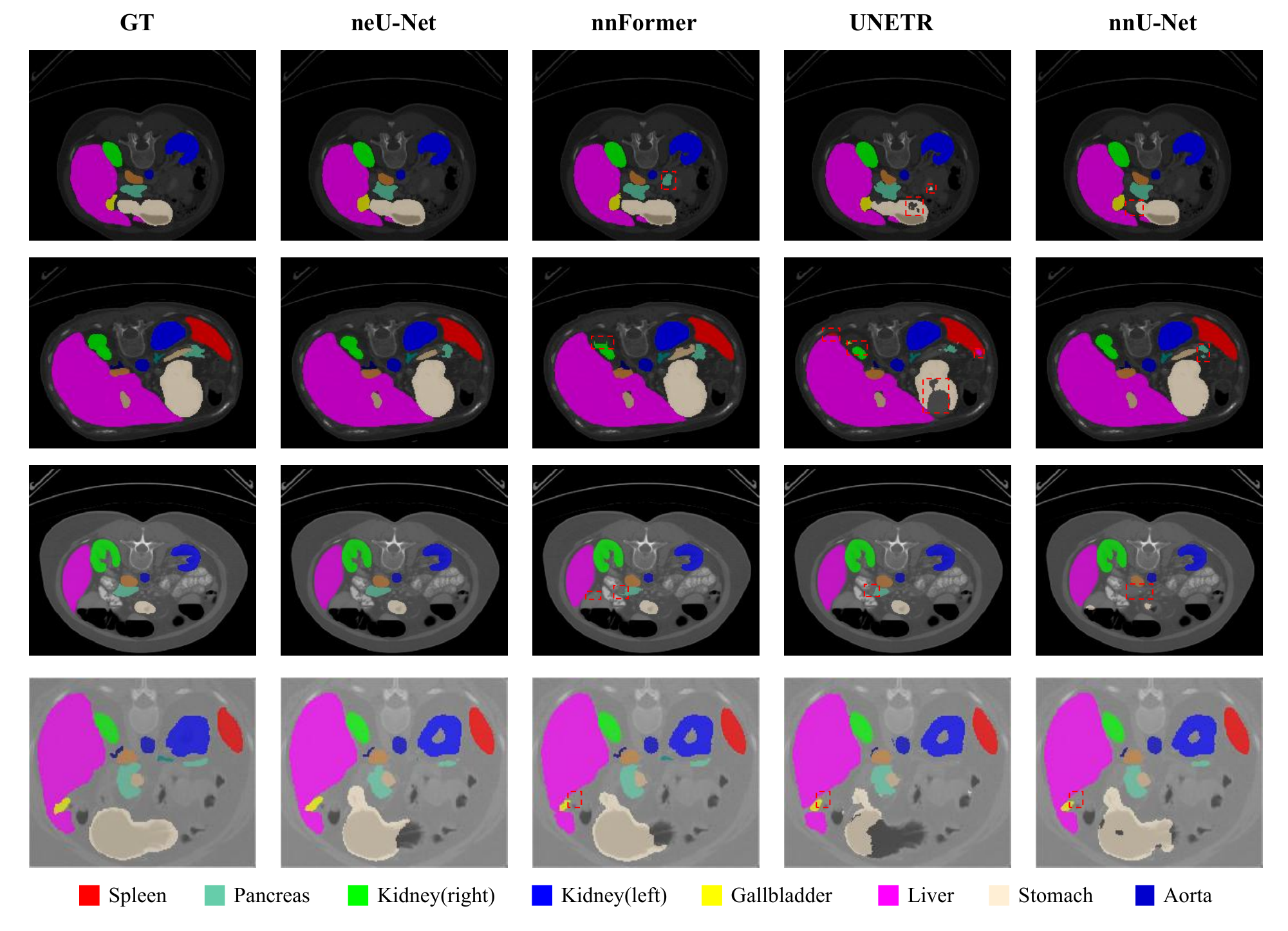}
 \caption{Qualitative comparison of different models in Synapse dataset. nnFormer and UNETR are methods based on the vision transformer\citep{vit}, while nnU-Net is a powerful medical image segmentation framework based on CNN. neU-Net significantly improves segmentation quality by introducing multi-scale wavelet inputs and sub-pixel convolution}
 \label{fig4}
\end{figure*}
\begin{figure}[!t]
    \centering
    \includegraphics[width=0.4\textwidth]{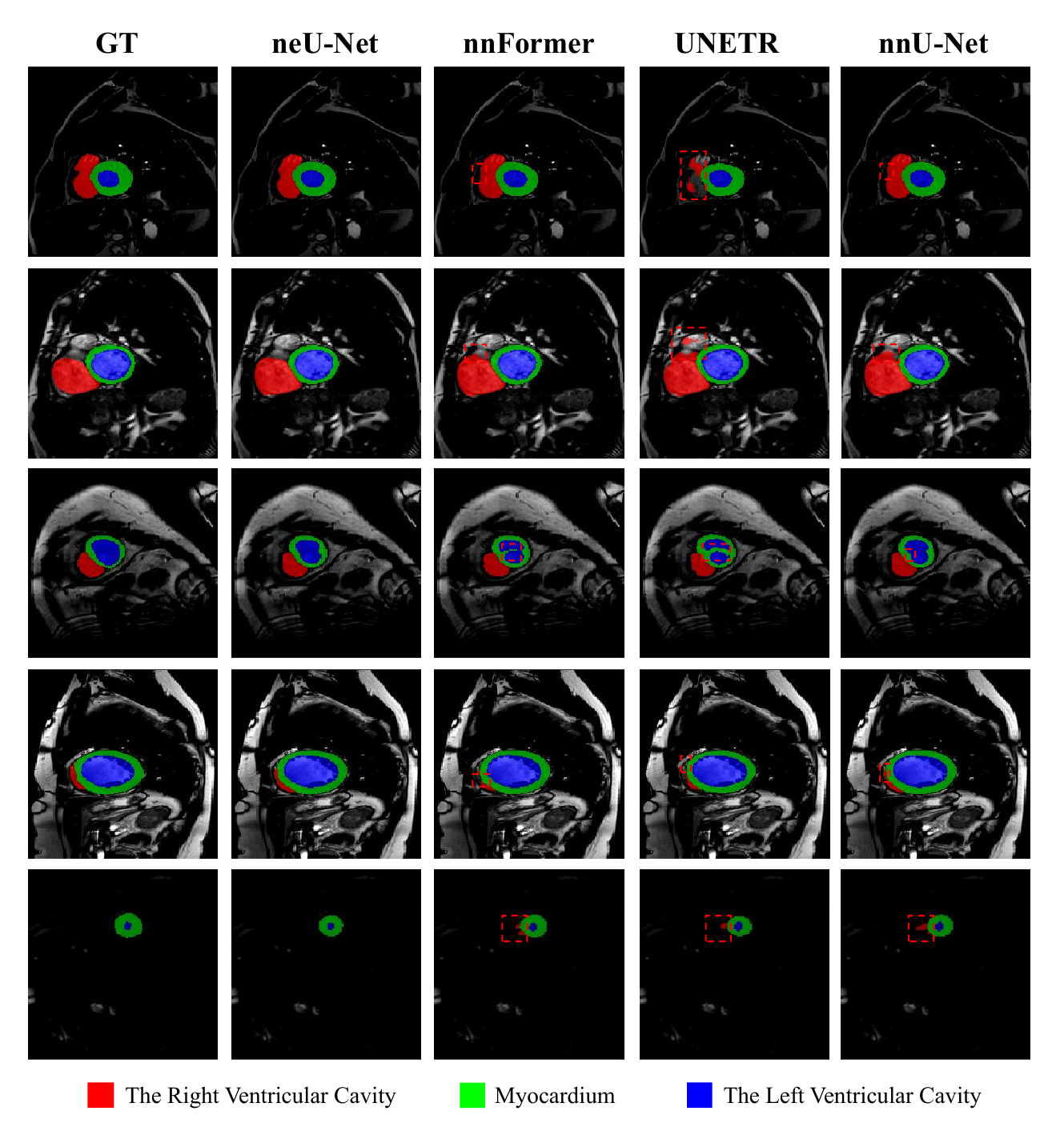}  
    \caption{Visualization of segmentation results on ACDC dataset}
    \label{figACDC}
\end{figure}
\subsection{Quantitative Results}
To validate the effectiveness of neU-Net on different segmentation tasks, we compared our model with other state-of-the-art methods on the Synapse and ACDC datasets. Table.\ref{table:synapse} shows the experimental results of all models on the multi-organ segmentation task. neU-Net achieved the highest average DSC and the lowest average HD95, reaching 87.83\% and 9.13mm, respectively. We significantly improved the segmentation performance of the right kidney, gallbladder, and aorta, with DSC improvements of 1.66\% , 9.13\% and 0.16\% , respectively, compared to second-best method nnU-Net.
Fig.\ref{fig4} illustrates the qualitative comparison between neU-Net and other methods on the Synapse dataset. As shown in the first row, our method improves the segmentation quality of the stomach and pancreas. In the second row, nnFormer and UNETR bothexhibit under-segmentation in the right kidney. Additionally, UNETR misses a substantial portion of the stomach and confuses the liver and spleen in some areas. The performance of nnU-Net on pancreas segmentation is not ideal. In contrast, neU-Net successfully delineates the boundaries of these organs. In the third row, neU-Net effectively avoids under-segmentation of the pancreas. The fourth row demonstrates the significant improvement of our method in gallbladder segmentation.

\begin{table}[h!]
        \centering
        \caption{Comparison with other models on ACDC dataset. We evaluate the performance of each model using DSC metric. Abbreviations stand for: RV: \textit{right ventricle}, LV: \textit{left ventricle} and Myo: \textit{myocardium}.
	}
        \resizebox{0.5\textwidth}{!}{%
	\begin{tabular}{l|ccc|c}
		\toprule
		Methods & RV & Myo & LV & Average  \\
		\midrule
		\midrule
		TransUNet~\citep{transunet} 
		& 88.86  & 84.54 & 95.73 & 89.71 \\
		Swin-UNet~\citep{swinunet}
		& 88.55  & 85.62 & 95.83 & 90.00 \\
		UNETR~\citep{unetr}
		& 85.29 & 86.52 & 94.02 & 86.61 \\
		MISSFormer~\citep{missformer} 
		& 86.36 & 85.75 & 91.59 & 87.90 \\
		nnFormer~\citep{nnformer} 
		& \textbf{90.94} & 89.58 & \textbf{95.69} & 92.06 \\
		nnU-Net~\citep{nnUNet} 
		& 90.24  &89.24  &95.36  &91.62 \\
		\midrule
		\rowcolor{red!6} neU-Net(Ours) & 90.75 & \textbf{89.91}& 95.66& \textbf{92.11} \\
		\bottomrule
	\end{tabular}}
	\vspace{0.1em}
	
	\label{table:ACDC}
	\vspace{-0.1cm}
\end{table}
\begin{table*}[!t]
        \centering
        \caption[caption]%
	{\upshape Leaderboard Dice coefficient ablation results for multi-organ segmentation in the BTCV challenge. Please note the abbreviations: Spl for spleen, RKid for right kidney, LKid for left kidney, Gall for gallbladder, Eso for esophagus, Liv for liver, Sto for stomach, Aor for aorta, IVC for inferior vena cava, Veins for portal and splenic veins, Pan for pancreas, and AG for left and right adrenal glands.}
	\scriptsize
	\resizebox{\textwidth}{!}{%
		\begin{tabular}{l|lll|cccccccccccc|c}
			\midrule \midrule
			Baseline &MWA &MW &SPC & \multicolumn{1}{l}{Spl}  
			& \multicolumn{1}{l}{RKid} & \multicolumn{1}{l}{LKid} 
			& \multicolumn{1}{l}{Gall}  & \multicolumn{1}{l}{Eso} 
			& \multicolumn{1}{l}{Liv} & \multicolumn{1}{l}{Sto} 
			& \multicolumn{1}{l}{Aor} & \multicolumn{1}{l}{IVC} 
			& \multicolumn{1}{l}{Veins}   & \multicolumn{1}{l}{Pan} 
			&\multicolumn{1}{l}{AG} & \multicolumn{1}{|l}{Avg.} \\ \hline
			\quad &\quad &\quad  &\quad     
			& 80.41 & \textbf{89.59}                   
			& 86.94 & 56.00                        
			& 73.17 & 90.49                        
			& 86.03 & 89.10   
			& 88.24 & 67.49
			& 70.67 & 65.53
			& 78.64 \\
			\multirow{2}{*}{nnU-Net~\citep{nnUNet}} &\quad &\quad  &\checkmark 
			& \textbf{80.47} & 87.89                        
			& 79.97 & \textbf{68.44}                        
			& 74.19 & 90.60                     
			& 82.88 & 90.89   
			& 87.70 & 70.00
			& 77.73 & 66.02
			& 79.73
			\\ 
			\quad &\checkmark  &\quad  &\checkmark 
			& 80.45 & 88.36                  
			& 81.55 & 56.52                   
			& 76.94 & \textbf{90.62}                     
			& \textbf{87.06} & 88.84  
			& 88.50 & 69.98
			& 77.37 & 66.52
			& 79.39
			\\ 
			&\quad &\checkmark &\checkmark          
			& 80.42 & 88.12                        
			& 81.39 & 67.62                        
			& \textbf{78.38} & 90.49                         
			& 86.41 & 90.34   
			& 87.81 & 69.42
			& 77.45 & 65.4
			& \textbf{80.28}   
			\\ \hline
		\end{tabular}%
	}
	\\
	\label{abaltion study}
\end{table*}

Table.\ref{table:ACDC} presents the experimental results on the ACDC dataset, our model achieved the best average DSC of 92.11\%. The segmentation performance of neU-Net on myocardium was improved by 0.33\% compared to the second-place nnFormer\citep{nnformer}. Fig.\ref{figACDC} illustrates the visualization results, in the first row, nnFormer and nnU-Net have oversegmentation issues for the right ventricular cavity, UNETR exhibits significant undersegmentation, while neU-Net effectively improves the segmentation of the right ventricular cavity. In the second and fifth rows, the other networks exhibit over-segmentation of the right ventricular cavity, while in the fourth row, the other methods result in under-segmentation. In contrast, our model effectively delineates its boundaries.
The results in the third row demonstrate the significant improvement of neU-Net in myocardium segmentation.

\subsection{Ablation study}

In this chapter, we conducted meticulous ablation experiments on the proposed modules, integrated into the nnU-Net framework \citep{nnUNet}, utilizing the BTCV dataset to rigorously evaluate their effectiveness. Our focus was on dissecting the roles of three pivotal modules: (1) Multi-scale Wavelet Downsampling of Approximation Coefficients (MWA) as inputs, (2) Multi-scale Wavelet Coefficients (MW) as inputs, and (3) Sub-pixel Convolution (SPC). The evaluation process revolved around the Dice Similarity Coefficient (DSC), a pivotal metric that gauges the likeness between predicted segmentations and the ground truth. By deliberately deactivating or adapting these modules while ensuring the constancy of other components, we systematically unveiled their individual contributions. The DSC metric, serving as a robust benchmark, effectively illuminated the distinctive impact of each module on segmentation performance. The detailed results are presented in Table.\ref{abaltion study}.

After the introduction of Sub-pixel Convolution (SPC), the Dice Similarity Coefficient (DSC) experienced an improvement of 1.09\%. As evident from the results in the Table.\ref{abaltion study}, there were substantial enhancements in the DSC scores of organs such as gallbladder, veins, and pancreas, which originally had lower DSC scores. These organs, characterized by their smaller volumes compared to others, exhibited significant improvements, highlighting the superior sensitivity of the SPC module towards smaller targets.\par

However, the addition of SPC alone led to a slight decrease in the segmentation results for larger targets. This outcome was not desirable. In order to address this phenomenon, the introduction of multi-scale wavelet approximation coefficients was proposed to complement the input of each encoding layer. From the experimental results, it can be observed that while there was a marginal overall decrease in the Dice Similarity Coefficient (DSC) by approximately 0.57\%, the segmentation performance for the kidneys showed a significant improvement.\par
\begin{figure}[t]
	\centering
	\includegraphics[width=0.4\textwidth]{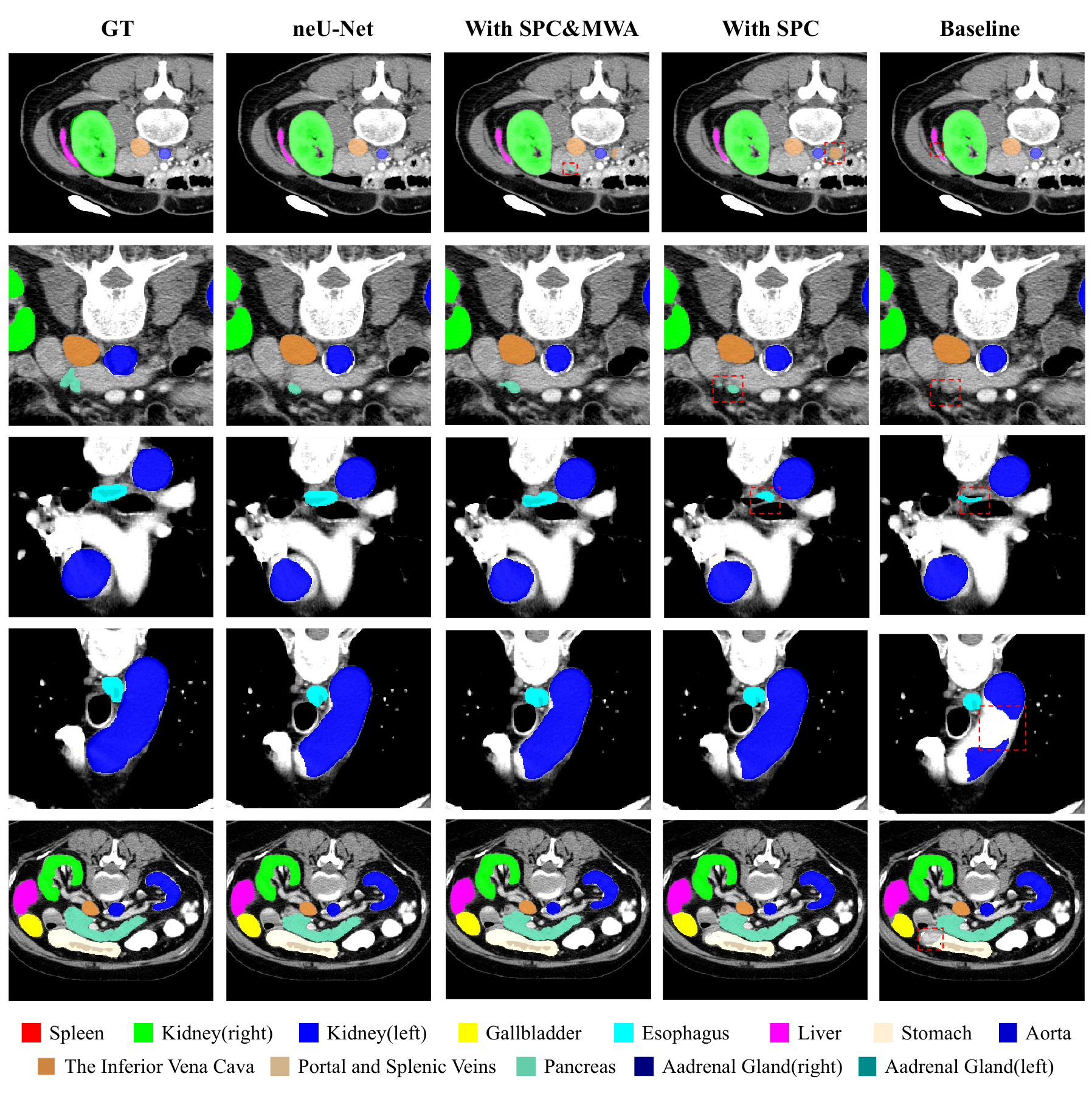}
	\caption{Quantitative ablation experiment results were compared on the BTCV dataset, utilizing the nnU-Net as the baseline method. For the sake of clarity, the introduction of the SPC module on the baseline method was denoted as With SPC. The addition of both the SPC and MWA modules was referred to as With SPC\&MWA, while the integration of the SPC and MW modules was labeled as neU-Net.}
	\label{fig5}
\end{figure}

Building upon the incorporation of MWA, we took a step further by introducing all wavelet coefficients, forming the MW module, which supplements high-frequency details. This combination with SPC's superior performance in small target segmentation was anticipated. Experimental results validated that with the inclusion of both the MW and SPC modules, the average Dice Similarity Coefficient (DSC) increased by 1.64\% compared to the original nnU-Net framework. Furthermore, the introduction of detail coefficients led to a 0.89\% improvement over the MWA module. Fig.\ref{fig5} provides a more intuitive visualization of these findings.Note that: For the sake of clarity, the introduction of the SPC module on the baseline method was denoted as With SPC. The addition of both the SPC and MWA modules was referred to as With SPC\&MWA, while the integration of the SPC and MW modules was labeled as neU-Net.

\section{Conclusion}
We have identified an imbalance in the evolution of commonly used encoder-decoder structures. While encoders have grown increasingly complex, decoders have often been overlooked. Furthermore, given the specific characteristics of medical image data, the complexity of encoders may not necessarily lead to optimal performance. Therefore, we have introduced two high-level strategies to enhance existing models: the incorporation of additional information and the development of superior decoders. \par
We have put these concepts to the test by introducing multi-scale wavelet transformation (MW) to supplement additional information and proposing the upsampling module SPC to enhance decoder performance within the nnU-Net framework. \par
From the experimental results, it is evident that our two primary ideas have been substantiated through comparisons with Current leading approaches. Our neU-Net has achieved new SOTA results on two datasets, Synapse and ACDC. This underscores the promising avenues for further research and development, with a focus on leveraging additional information and refining decoder architectures for improved outcomes.
\section{Acknowledge}
This work was supported by National Natural Science Foundation of China under Grant 61301253 and the Major Scientific and Technological Innovation Project in Shandong Province under Grant 2021CXG010506 and 2022CXG010504;"New Universities 20 items" Funding Project of Jinan under Grant 2021GXRC108 and 2021GXRC024.
\bibliographystyle{model2-names.bst}\biboptions{authoryear}
\bibliography{paper}
\end{document}